\documentclass[a4paper,11pt]{article}
\pdfoutput=1 
\usepackage{graphicx,wrapfig} 
\usepackage{mwe}
\usepackage{subcaption} 
\usepackage{jinstpub} 

\title{\boldmath A compact cosmic muon veto detector and possible use with the Iron Calorimeter detector for neutrinos}

\author[a,b]{Neha,}
\author[a,b,1]{S. Mohanraj,\note{presently at Linnaeus University, Sweden}}
\author[a,b]{A. Kumar,}
\author[a,b]{T. Dey,}
\author[b]{G. Majumder,}
\author[b]{R. Shinde,}
\author[b]{P. Verma,}
\author[b]{\\B. Satyanarayana}
\author {and}
\author[b]{V.M. Datar}

\affiliation[a]{Homi Bhabha National Institute, Anushaktinagar, Mumbai-400094, India}
\affiliation[b]{Tata Institute of Fundamental Research, Homi Bhabha Road, Colaba, Mumbai-400005, India}

\emailAdd{neha$\_$005@tifr.res.in}

\abstract{The motivation for a cosmic muon veto (CMV) detector is to explore the possibility of locating the proposed large Iron Calorimeter (ICAL) detector at the India based Neutrino Observatory (INO) at a shallow depth. An initial effort in that direction, through the assembly and testing of a $\sim$1~m $\times$ 1~m $\times$ 0.3~m plastic scintillator based detector, is described.  The plan for making a CMV detector for a smaller prototype mini-ICAL is also outlined.}


\begin{document}
\maketitle
\flushbottom

\section{Introduction}
\label{sec:intro}

The proposed Iron Calorimeter (ICAL) detector is a 51 kiloton device consisting of 3 modules to study atmospheric neutrinos and is planned to be set up in the underground INO laboratory at Pottipuram, Tamil Nadu in India \citep{kumar2017}. The rock overburden is greater than 1~km in all directions and will reduce the cosmic ray background, mainly due to muons, by about a factor of $10{^6}$. On the other hand there are many advantages, as mentioned later, of having a large detector such as ICAL at a shallow depth. A rock overburden of 100~m reduces the hadronic and electromagnetic components of the cosmic ray background drastically. The cosmic ray muon induced events in ICAL could be identified, and eliminated in the analysis, by using an efficient cosmic muon veto (CMV) detector surrounding ICAL. However, muon induced hadronic and electromagnetic component could potentially pose a problem. Fortunately, as discussed later, it can be estimated and turns out to be negligible.

For such a Shallow depth ICAL (SICAL), it is clear that events due to cosmic muons need to be identified very efficiently and discarded. While the outer parts of the ICAL detector could be used for this purpose it might be more efficient and versatile if one builds an efficient muon veto detector to identify and reject cosmic muon events. Such a detector could, if suitably configured, be also useful for magnetic monopole searches \citep{nitali2015}, investigation of the KGF anomalous events  \citep{narasimham2004}, \citep{nitali2016} and perhaps in increasing the fiducial volume of ICAL. Another benefit of a SICAL would be the in-situ checks on the performance of the active detector elements using cosmic muons as also their possible use in getting information about the internal magnetic field in the iron via the Muon Spin Rotation technique \citep{neha_DAE-HEP}. If a shallow depth ICAL is feasible, one could envisage much larger iron calorimeter detectors for neutrino science as well as applications such as earth tomography. It may be pointed out that while the present study was being carried out it came to our notice that a sufficiently large CMV system had already been designed for the Mu2e experiment \citep{cdr_mu2e} at Fermilab.

\section{A small sized muon trigger and cosmic muon veto detector}
A modest step in the direction of building a proof-of-principle CMV detector was to assemble a small sized cosmic muon veto shield at Tata Institute of Fundamental Research using 14 plastic scintillator \citep{bicron} paddles of size 320~mm (width) $\times$ 960~mm (length) $\times$ 10~mm (thickness). Each of these had grooves about 4~mm wide and 2~mm deep at a pitch of 30~mm along the length for housing four 1~mm diameter Wave Length Shifting (WLS) fibres \citep{kuraray}. The 44 WLS fibres in 11 grooves were bunched together in a "cookie" and coupled  to a 50~mm diameter photomultiplier tube~(PMT). These paddles were arranged in the form of an inner and outer cosmic veto $"$hut$"$ each with 4 $"$walls$"$ consisting of 4 paddles standing on the long edge forming a footprint of a square of inner dimension about 920~mm and 3 paddles forming the $"$ceiling$"$. The top of the outer hut had paddles placed in a direction perpendicular to the ones in the inner hut $"$ceiling$"$ as shown in Fig.~1(a).  

In the first set of runs the cosmic muons were detected in a set of 4 plastic scintillators (PS) each of transverse dimensions 440~mm $\times$ 440~mm and 10~mm thick (MuL). Two sets of 2 PSs, each placed on top of one another, were separated by about $\sim$30~mm of lead as shown in Fig.~1(c), to reduce $\gamma$-induced coincidences.

\begin{figure*}[h]
    \centering
    \begin{subfigure}[t]{0.5\textwidth}
        \centering
        \includegraphics[width=1.05\linewidth]{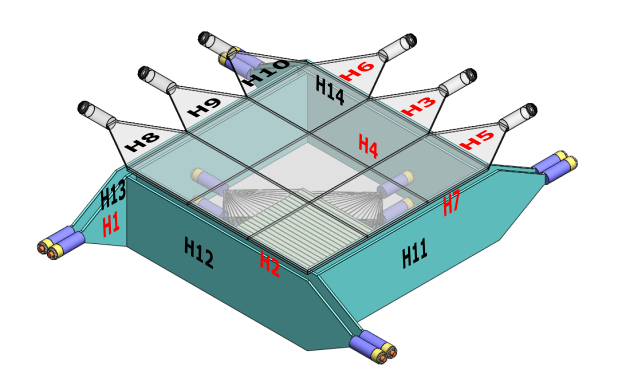}
        \caption{}
    \end{subfigure}%
        \begin{subfigure}[t]{0.5\textwidth}
       \centering
        \includegraphics[width=0.95\linewidth]{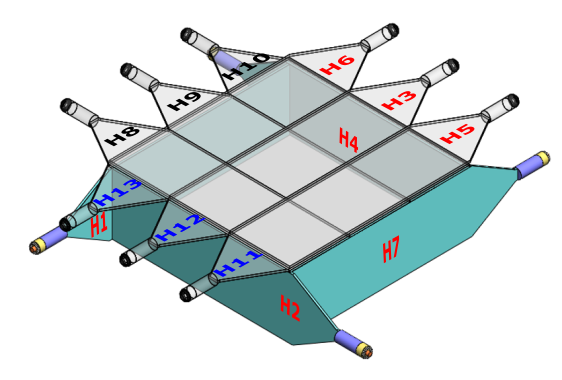}
        \caption{}
    \end{subfigure}%
        \newline
        \begin{subfigure}{0.5\textwidth}
        \centering
        \includegraphics[width=1\linewidth]{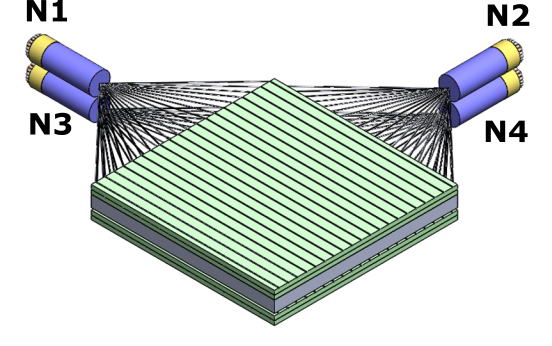}
        \caption{}
    \end{subfigure}%
    \begin{minipage}{0.5\textwidth}
        \centering
        \includegraphics[width=1\linewidth]{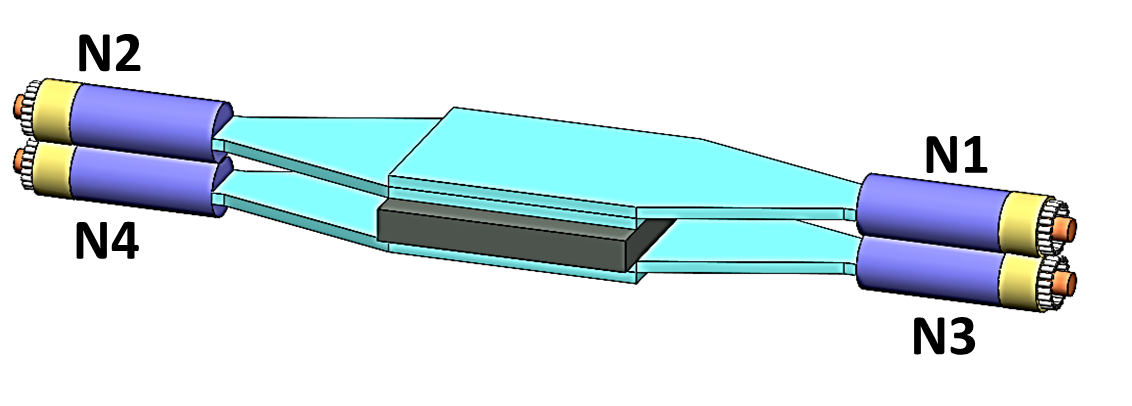}
        \subcaption{}
        \includegraphics[width=1\linewidth]{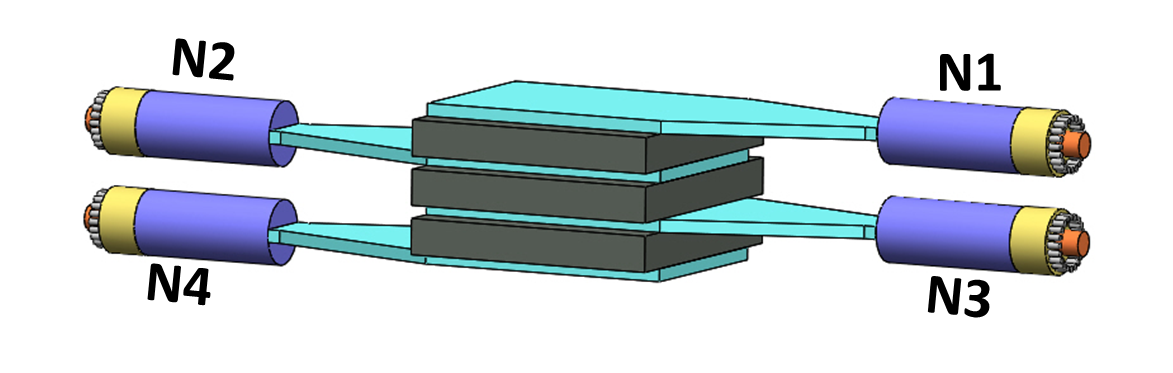}
        \subcaption{}
    \end{minipage}
    \caption{3D model of trigger detector configurations and CMV hut. (a) CMV shield with 2 layers of scintillators on 4 walls and top with inner hut paddles (H1-H7) labeled in red and outer hut paddles (H8-H14) in black (b) CMV shield with 3 layers of scintillators on the top, with topmost layer paddles (H11,H12,H13) labeled in blue, middle layer paddles (H8,H9,H10) in black and bottom layer paddles (H3,H5,H6) in red and 1 layer of scintillators on the 4 walls labeled in red (c) MuL (d) MuS (e) MuS along with 3 Pb layers.}
     \end{figure*}

\begin{figure}[h]
  \centering
    \includegraphics[width=0.5\textwidth]{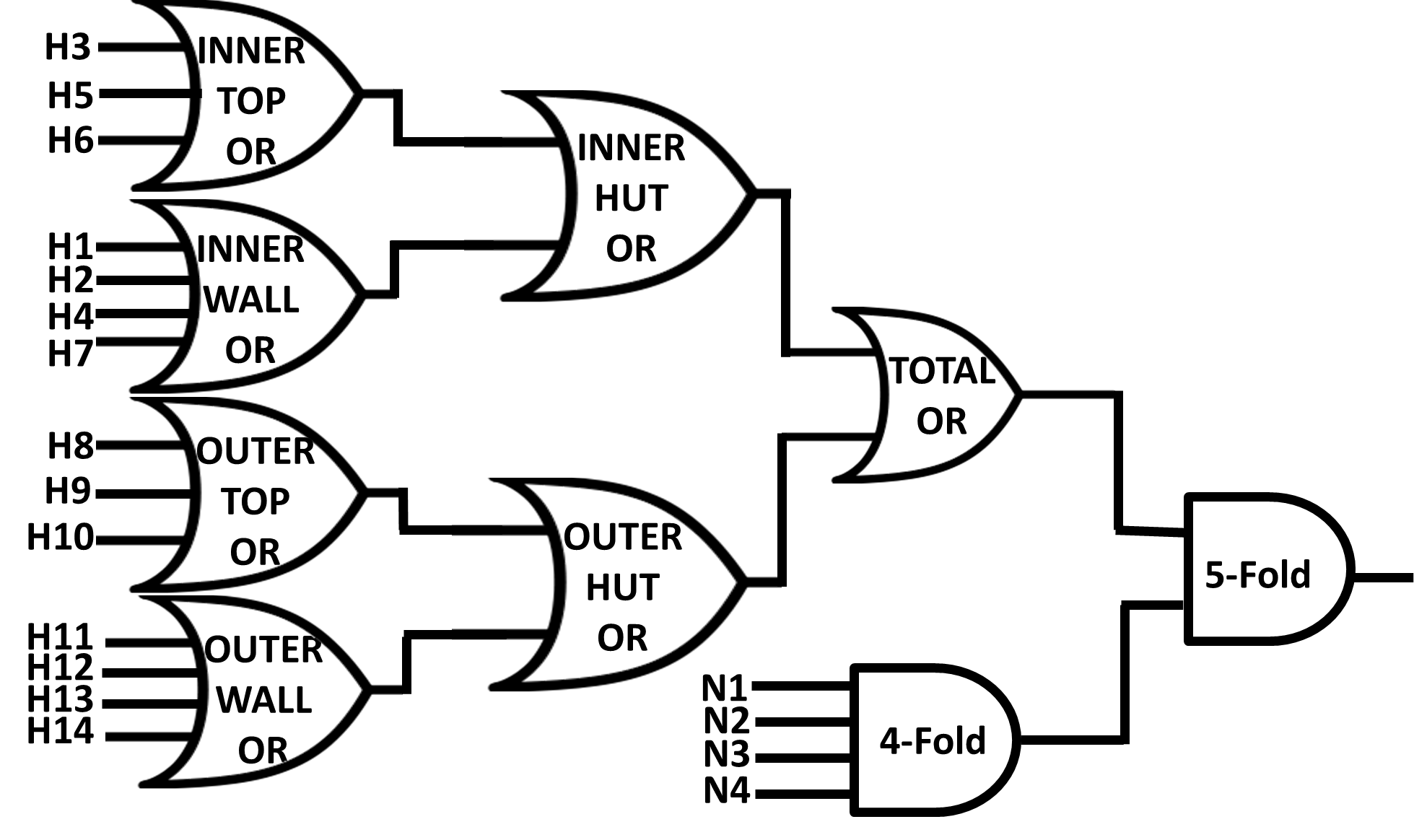}
  \caption{Logic diagram of Veto, 5-fold and 4-fold coincidence circuits.}
\end{figure}
A 4-fold coincidence ensured that the trigger generated was mainly due to cosmic muons. One of the PS (N1) fast logic signal was made narrow ($\sim$10~ns width) and delayed by about 20~ns while the other 3 had a width of about 50~ns before it is fed to the trigger generating 4-fold coincidence unit. This reduces the contribution of jitter due to N2, N3 and N4 to the timing data. The muon trigger rate  was $\sim$15~Hz for MuL configuration. The logic signals of the hut PSs were combined to make different OR signals as shown in Fig.~2 and their corresponding rates are mentioned in Table~1. The OR of 14 scintillators was then ANDed with the muon trigger to form a 5-fold which was used for efficiency calculations. Fast signals from each of the scintillators were also sent to 2 TDCs for measuring the relative times with respect to the trigger. 

\begin{table}[h]
\centering
\caption{\label{tab:i} Various OR rates.}
\smallskip
\begin{tabular}{|c|c|}
\hline
Description & Rates (Hz)\\
\hline
Inner Wall & 2350\\
Inner Top & 540\\
Inner Hut & 2830\\
Outer Wall & 3450\\
Outer Top & 2500\\
Outer Hut & 5900\\
\hline
\end{tabular}
\end{table}

Another configuration that was studied used a smaller sized muon trigger with a set of 4 plastic scintillator paddles of transverse size 150~mm $\times$ 200~mm and 10~mm thick (MuS) with $\sim$30~mm of lead layer in between N2 and N3 as shown in Fig.~1(d). These could be placed more centrally with respect to the 920~mm $\times$ 920~mm space available inside the 2 layer CMV hut as shown in Fig.~3(b). The muon trigger rate in this case was $\sim$2~Hz. 
 
In order to study the inefficiency of the CMV detector, with the MuS trigger detector, a third layer was added on top of the upper two layers. Three side paddles (H11, H12, H13) from the outer hut were used for this purpose as shown in Fig.~1(b). The third layer of paddles on top was placed parallel to, and with an offset of 20~mm, from the 2nd layer of paddles so as to cover the gaps between the adjoining scintillators.

To study the effect of lead in reducing coincidences triggered by $\gamma$-ray, a fourth configuration with lead layers in between consecutive scintillators of MuS trigger detector was assembled as shown in Fig.~1(e). The description of all the configurations of hut and trigger scintillators is given in Table 2. 

\begin{figure}[h]
\centering 
\includegraphics[width=0.47\textwidth,origin=l,angle=0]{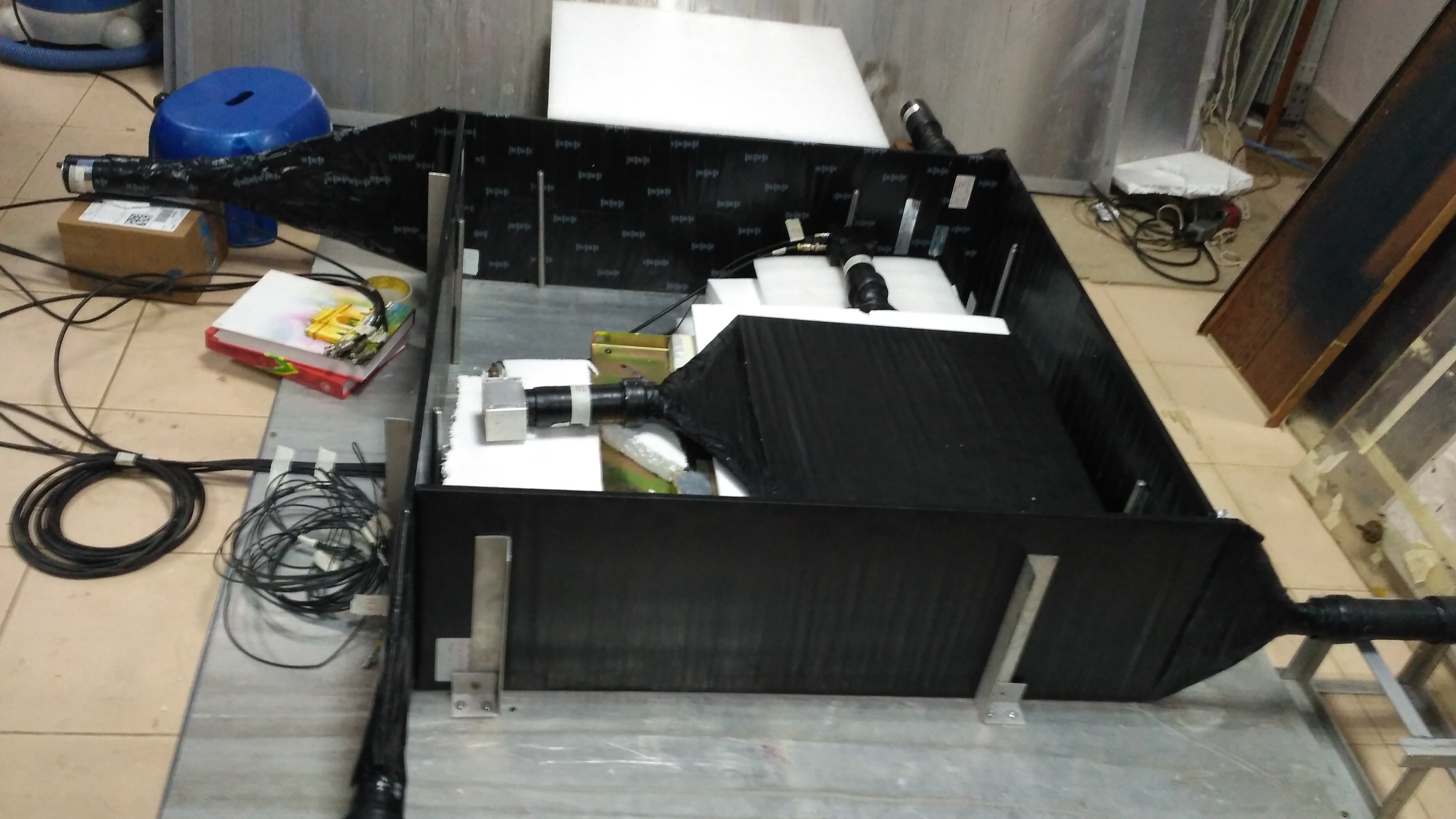}
\qquad
\includegraphics[width=0.47\textwidth,origin=r,angle=0]{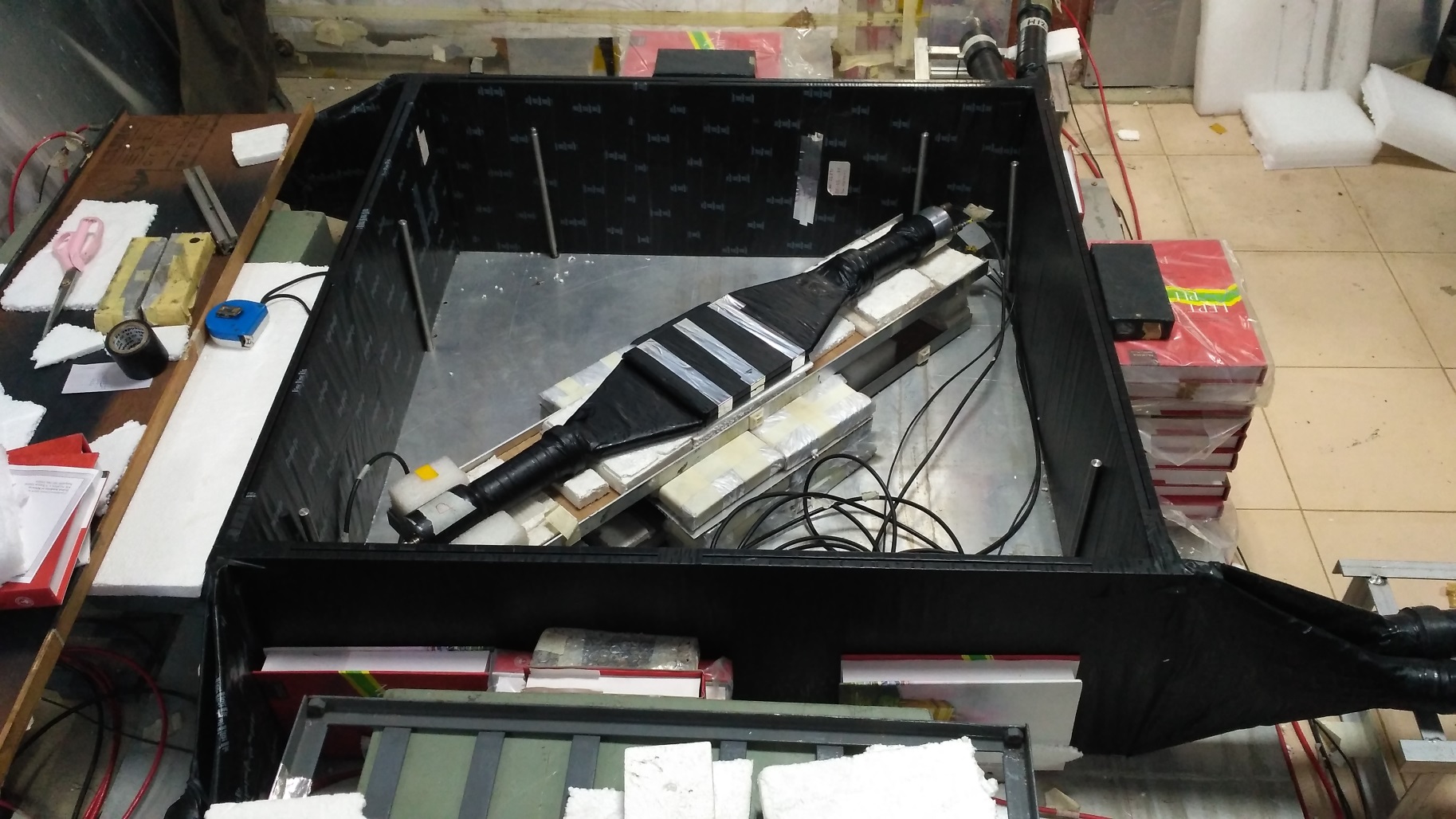}
\caption{\label{fig:i} Photographs of the MuL (left) and MuS (right) trigger detectors inside the CMV detector with the top layers removed.}
\end{figure}

\begin{table}[h]
\centering
\caption{\label{tab:i} Description of all detector configurations.}
\smallskip
\begin{tabular}{|c|c|c|c|c|c|}
\hline
Configuration & No. of scintillator & No. of scintillator & Trigger  & No. of Pb & Figure \\
 & layers on the top in & layers at the sides in & detector & layers & No.\\
 & CMV detector & CMV detector & & & \\
\hline
I & 2 & 2 & MuL & 1 & 1(a,c)\\
II & 2 & 2 & MuS & 1 & 1(a,d)\\
III & 3 & 1 & MuS & 1 & 1(b,d)\\
IV & 3 & 1 & MuS & 3 & 1(b,e)\\
\hline
\end{tabular}
\end{table}

The time resolution was measured to be $\sim$1~ns for trigger scintillators in the MuS configuration and $\sim$2~ns for top hut scintillators. Typical time spectra are shown in Fig.~4. The photon transport within the scintillator results in a spread in timing. The hut scintillators are bigger in size than the trigger scintillators due to which the path-length transversed by the photon to reach the PMT is more. This results in a larger time resolution for the hut scintillators.  

\begin{figure}[h]
\centering 
\includegraphics[width=1.0\textwidth,origin=c,angle=0]{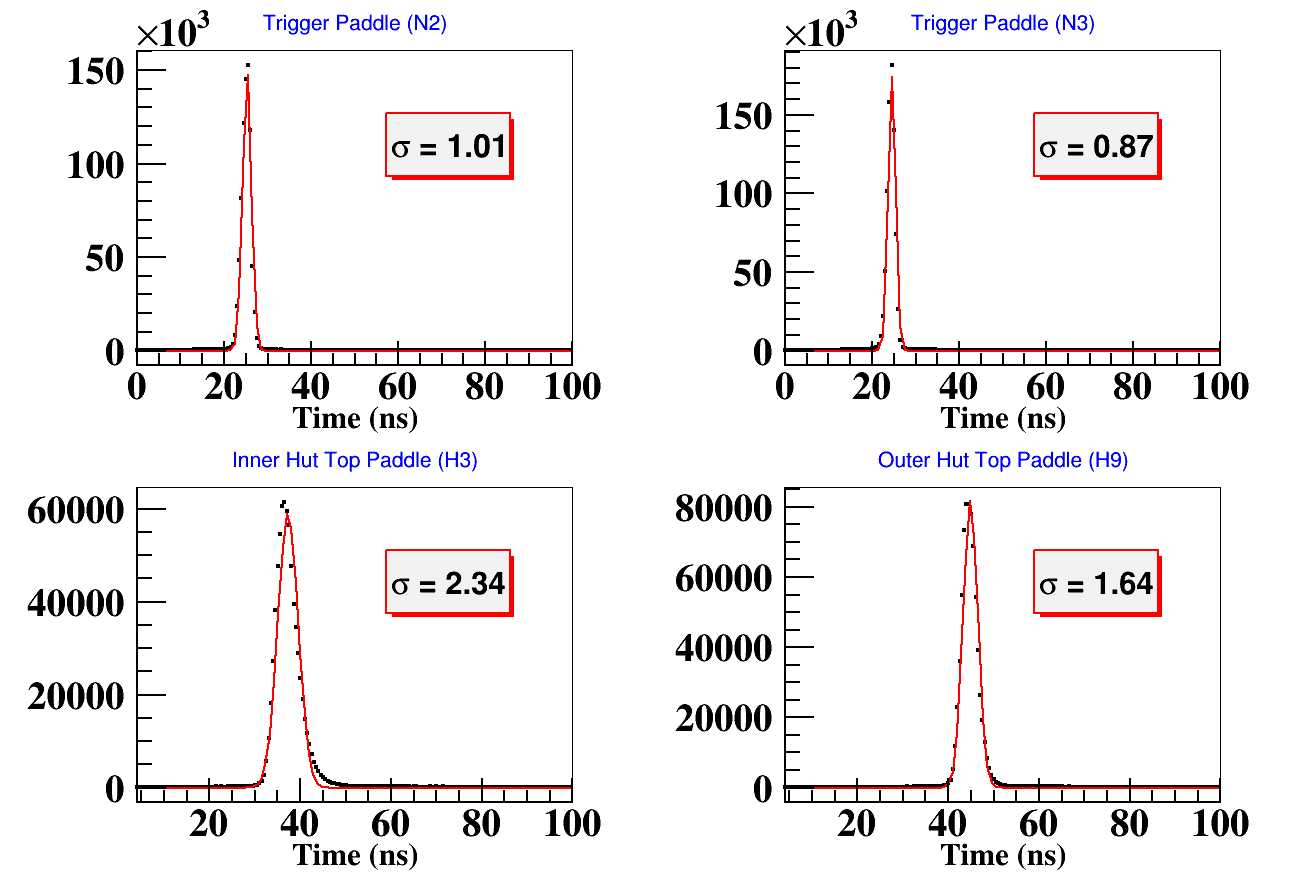}
\caption{\label{fig:i} TDC Spectra of trigger and CMV hut scintillators.}
\end{figure}

\section{Results and discussion}
\subsection{Cosmic Muon Rejection Efficiency}

The efficiency for the rejection of cosmic muons from the ratio of the 5-fold to 4-fold coincidences was calculated after analyzing the event-by-event TDC data in a CAMAC based data acquisition system.  Table~3 show the veto efficiencies for individual huts, inner and outer, as well as for the combined huts (ORed) for configurations I and II. The TDC was operated at a full scale range of 100~ns. 

\begin{table}[h]
\centering
\caption{\label{tab:i} CMV efficiencies for individual parts of the detector at -20~mV threshold for configuration I and II.}
\smallskip
\begin{tabular}{|c|c|c|}
\hline
Veto configuration&Veto efficiency ($\%$)&Veto efficiency ($\%$)\\
 & (Configuration I) & (Configuration II)\\

\hline
Only Inner hut top & 91.347 $\pm$ 0.057 & 94.944 $\pm$ 0.026\\
Only Inner hut side & 1.111 $\pm$ 0.021 & 0.085 $\pm$ 0.003\\
Inner hut top and side & 6.972 $\pm$ 0.052 & 4.670 $\pm$  0.025\\
Inner hut all & 99.420 $\pm$  0.015 & 99.680 $\pm$ 0.007\\
Only Outer hut top & 93.026 $\pm$ 0.052 & 95.652 $\pm$ 0.024\\
Only Outer hut side & 1.344 $\pm$ 0.023 & 0.099 $\pm$ 0.004\\
Outer hut top and side & 4.632 $\pm$ 0.043 & 3.935 $\pm$ 0.023\\
Outer hut all & 99.354 $\pm$ 0.016 & 99.652 $\pm$ 0.007\\
Inner + Outer huts & 99.880 $\pm$ 0.007 & 99.847 $\pm$ 0.005\\
\hline
\end{tabular}
\end{table}

The analysis of the event-by-event TDC data gives an overall veto efficiency of (99.880 $\pm$ 0.007)\% for the configuration I. The inner and outer huts have individual efficiencies of (99.420  $\pm$  0.015)\% and (99.354 $\pm$ 0.016)\%, respectively. As might be expected from the geometry, the side detectors contribute a smaller, but important, veto efficiency with the nearer two sides contributing more than the farther two.  The configuration II data gives an overall veto efficiency of (99.847 $\pm$ 0.005)\% with the inner and outer hut efficiencies being (99.680 $\pm$ 0.007)\% and (99.652 $\pm$ 0.007)\%, respectively. It must be mentioned that the 4-fold coincidence in the muon trigger detector is assumed to arise from cosmic muons. If a small fraction of these is due to high energy neutral particle cascades generated in the topmost muon trigger detector or the lowermost part of the top veto detector, they may not generate a signal in the CMV hut and could masquerade as a contribution to the inefficiency.
The TDC information of each paddle was analyzed to extract the 5-fold contributions due to different parts of the CMV detector. In configuration II of the detector assembly, geometrically no muon trajectory is possible which can generate a 5-fold due to contribution (a)  only from the side scintillators or (b) from both side and top scintillators. But 0.1$\%$ of the total 5-folds were observed due to (a) and 4.5$\%$ due to (b). This could be a result of simultaneous arrival of many secondary or tertiary particles of cosmic showers. 

For configuration III, a set of measurements was performed by varying the thresholds of the MuS trigger paddles from -20~mV to -150~mV to emphasize muons and reduce the electromagnetic shower contribution. The results from these runs are shown in Table~4. An increased threshold results in increase in veto efficiency from (99.823 $\pm$ 0.012)\% to (99.922 $\pm$ 0.010)\%. A larger threshold reduces the chance of trigger generation inside the hut due to low energy particles which may not give a hit in the CMV hut paddles resulting in inefficiency. This also reduces the rate of random coincidence of MuS trigger.

\begin{table}[h]
\centering
\caption{\label{tab:i} Veto efficiencies for different thresholds of configuration III.}
\smallskip
\begin{tabular}{|c|c|c|}
\hline
Threshold (mV) & Veto efficiency ($\%$)\\
\hline
-20 & 99.823 $\pm$ 0.012\\
-50 & 99.887 $\pm$ 0.011\\
-80 & 99.880 $\pm$ 0.011\\
-100 & 99.922 $\pm$ 0.010\\
-120 & 99.890 $\pm$ 0.014\\
-150 & 99.888 $\pm$ 0.013\\
\hline
\end{tabular}
\end{table}

\subsection{$\gamma$-ray induced mis-identification Rate}

The effect of removing or placing lead between successive layers of the plastic scintillators constituting the muon trigger MuS on the CMV detector efficiency was studied. 
With no lead in MuS for configuration III, the veto efficiency dropped to (99.16 $\pm$ 0.01)$\%$. This is significantly smaller than the efficiency with one layer of lead in configuration II, as seen from Table~3 strongly suggesting that $\gamma$-ray induced events are being misidentified as muons leading to a lower CMV efficiency. To test this hypothesis, configuration IV with three layers of lead were placed in between successive plastic scintillators of MuS, as shown in Fig.~1(e), was studied . The results are summarized in the Table~5.
\begin{table}[h!]
\centering
\caption{\label{tab:i} Veto efficiencies of configuration IV at -100~mV threshold for the trigger detectors.}
\smallskip
\begin{tabular}{|c|c|c|}
\hline
Description & Veto efficiency($\%$) & Veto efficiency($\%$)\\
 & (with 100~ns prompt gate) & (from fit)\\

\hline
Top Layer & 99.749 $\pm$ 0.006 & 99.52 $\pm$ 0.010\\
Middle layer & 99.951 $\pm$ 0.002 & 99.920 $\pm$ 0.005\\
Bottom Layer & 99.933 $\pm$ 0.003  & 99.908 $\pm$ 0.006\\
Top or middle Layers & 99.970 $\pm$ 0.002 & 99.964 $\pm$ 0.004\\
Top or bottom Layer & 99.973  $\pm$ 0.002 & 99.969 $\pm$ 0.004\\
Middle or bottom Layer & 99.979 $\pm$ 0.002 & 99.975 $\pm$ 0.003\\
Any one Layer & 99.981 $\pm$ 0.002 & 99.978 $\pm$ 0.003\\
\hline
\end{tabular}
\end{table}

The Veto efficiency is calculated by two ways:
\begin{enumerate}
\item Selecting events only from a prompt time window of 100~ns in the TDC data \item Selecting events in different time windows starting from 40~ns to 80~ns in steps of 10~ns.
\end{enumerate}
These efficiencies are then plotted as a function of time window and fitted to a polynomial of order one. This is being done in order to remove the contribution of chance coincidence to the efficiency calculations. The veto efficiency is the value where the fitted line cuts the y-axis i.e., the intercept of the curve. The typical plot is shown in Fig.~5 for the configuration where all three layers are used for the veto.

\begin{figure}[h]
\centering 
\includegraphics[width=0.7\textwidth,origin=c,angle=0]{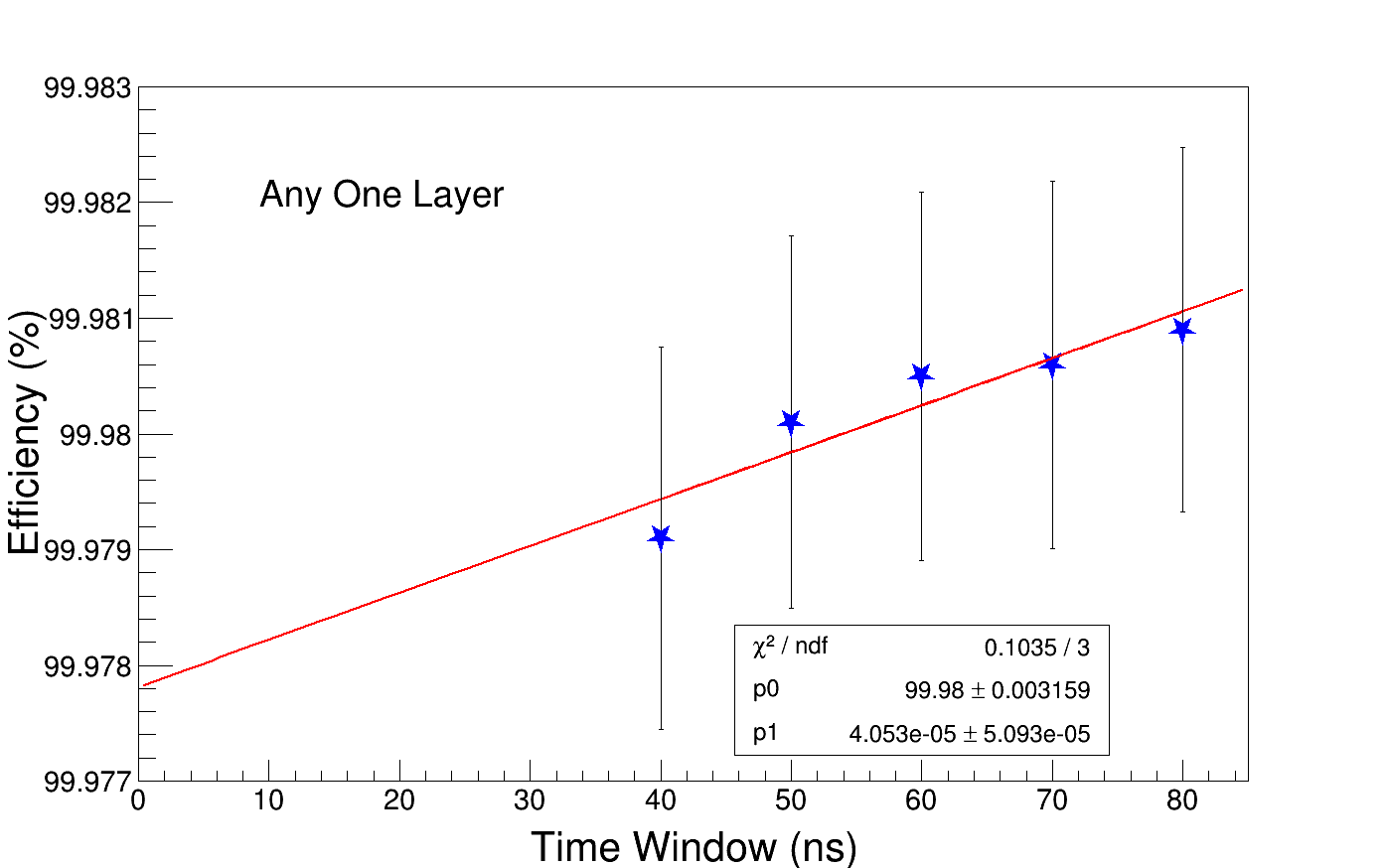}
\caption{\label{fig:i} Veto Efficiency as a function of time window.}
\end{figure}

\subsection{Other sources of inefficiencies in CMV}

In order to investigate the source of inefficiency of the CMV, the plastic paddles used in constructing the CMV huts were placed in the 12 layer Resistive Plate Chamber ($\sim$ 1~m $\times$ 1~m) stack detecting cosmic muons \citep{bhuyan2012} to measure the local inefficiency for muon detection as a function of X-Y position. Each RPC has 32 strips of width 28~mm and pitch of 30~mm for both X and Y sides. The muon trigger was generated by the coincidence of layers 4, 5, 6 and 7 with a trigger rate of $\sim$50~Hz. Hence, the efficiency for a 320~mm $\times$ 320~mm pixel can be measured to about 0.17\% accuracy in a day.

\begin{figure*}[h]
   \centering
    \begin{subfigure}[t]{0.45\textwidth}
    \centering
        \includegraphics[width=1\linewidth]{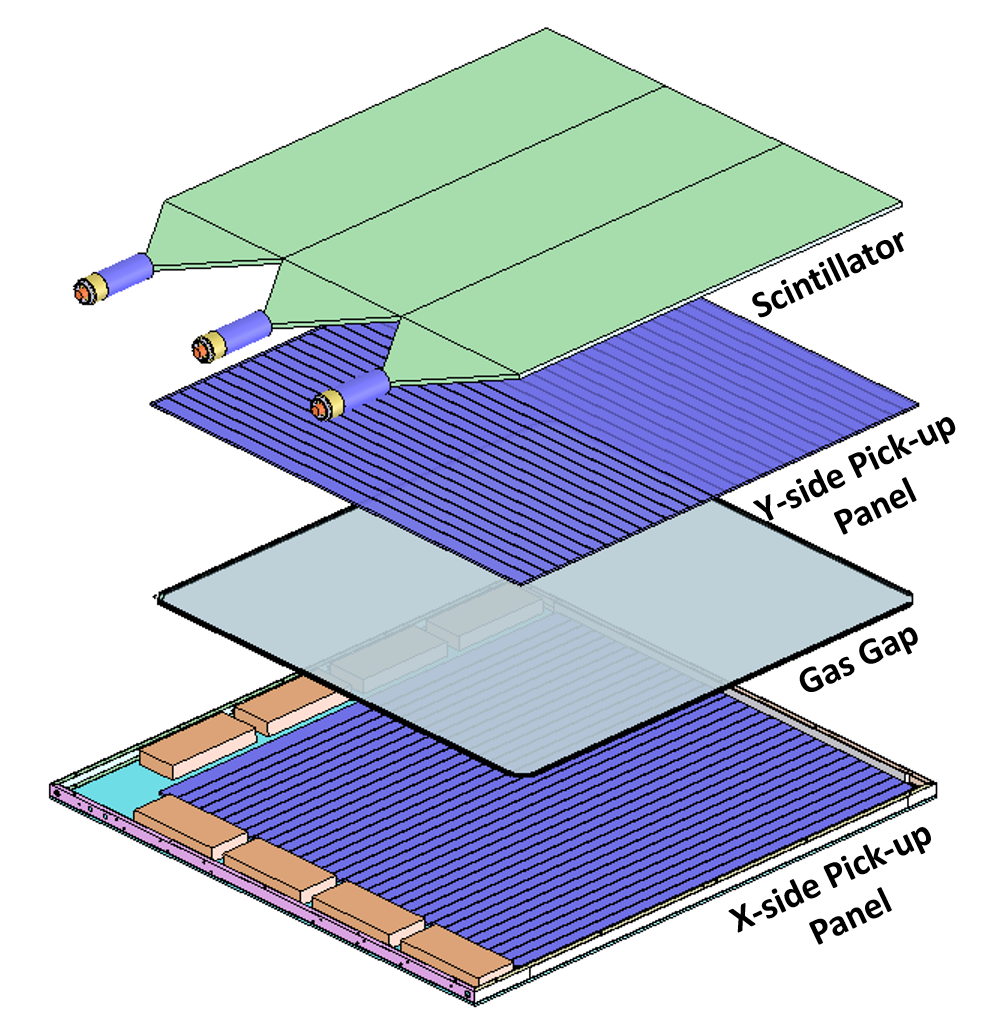}
        \caption{}
    \end{subfigure}%
        \begin{subfigure}[t]{0.45\textwidth}
        \centering
        \includegraphics[width=0.7\linewidth]{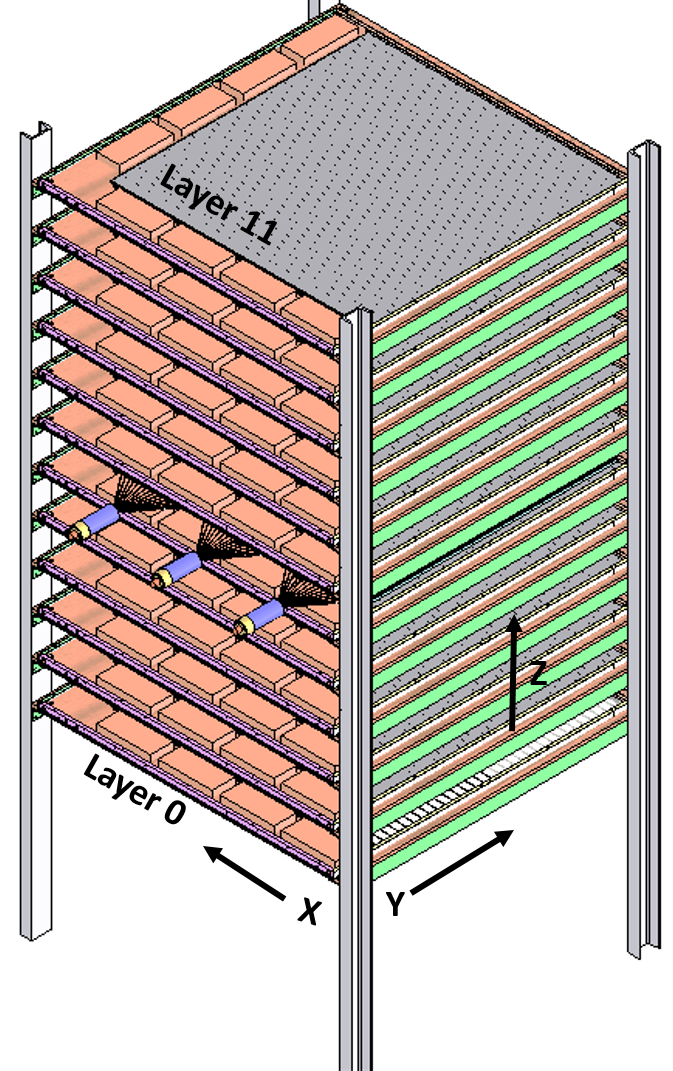}
        \caption{}
    \end{subfigure}%
    \caption{(a) Exploded view showing placement of 3 scintillators on top of 5th layer RPC with pick-up panels shown in blue color and gas gap in grey color, (b) the positioning of scintillators in the RPC stack.}
\end{figure*}

Three paddles were placed side by side on top of the 5th layer which ensures that the muon has passed through the scintillators as shown in Fig.~6(b). The total area covered by the paddles was 960~mm $\times$ 960~mm, which is effectively the active area of RPC. The hit information of the paddles, along with the hit information of RPCs, was stored in the latch data of the Data Acquisition system of RPC stack \citep{behere2013} for every event. The timing information of the hit was stored in commercial multi-hit TDC \citep{caentdc}. The X and Y positions were extracted from the raw RPC data using a straight line fit for the muon trajectory and based on the available TDC information, we estimate the inefficiencies in 3 zones (320~mm $\times$ 320~mm) of single paddle, which were plotted and shown in Fig.~7. This measurement was done after taking data in configurations I and II. The H6 paddle was found to be most inefficient of all the paddles as can be seen from Fig.~7. Replacing the PMT of H6 paddle resulted in improving the efficiency. It should be mentioned that the inefficiency of H6 did not contribute significantly to the veto efficiency as H6 was placed at the top-side of the CMV detector. The measurements in configurations III and IV were performed with the replaced PMT in H6 paddle.

\begin{figure}[h]
\centering 
\includegraphics[width=1.0\textwidth,origin=c,angle=0]{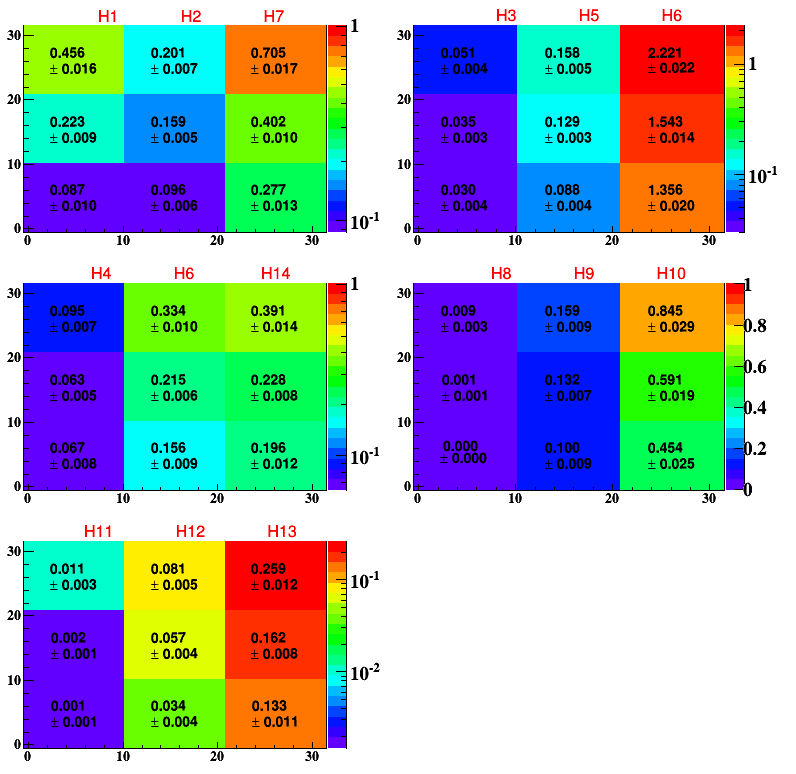}
\caption{\label{fig:i} The inefficiency map of all hut scintillators. H6 was most inefficient (topmost right). Middle-left shows the inefficiency map of H6 after replacing the PMT.}
\end{figure}

\subsection{Estimation of $\mu$-induced neutral background}

As mentioned in the introduction a potential source of background which can slip through the CMV undetected is that due to neutral hadrons or photons produced by the cosmic muons in the rock cover near the ICAL detector. Assuming a $\mu$ flux, ${\phi}$ and assuming a deep inelastic cross section producing these neutrals of 1~${\mu}$barn, their rate of production is given as

\begin{equation} 
\begin{split}
\frac{Rate}{\phi} & = \sigma n  = \frac{\sigma N_{A} L_{Had}}{A} \approx 10^{-6} 
\end{split}
\end{equation}
\\
where $L_{Had}$ $\sim$ 100~gm cm$^{-2}$ is the nuclear interaction length in rock (assumed to be SiO$_{2}$), N$_{A}$ is Avagadro's number, A is corresponding to 1~gm mole of SiO$_{2}$. In the above calculation if $\phi$ is assumed to be attenuated by a factor of 100 because of 100~m depth in the rock. In addition if the accompanying deep inelastically scattered (DIS) muon enters the CMV the hadron induced events can be identified and vetoed. Of the neutral hadrons produced the $\pi{^0}$ decays quickly and need not be considered further. The neutral long lived  kaon, ${K_L}^0$, events where the DIS muon goes undetected can contribute to hadron induced background in ICAL. However the average number of ${K_L}^0$ neutral kaons per muon DIS is expected to be smaller than that of pions. When the ${K_L}^0$ is accompanied by other charged particles, such as $\pi^{\pm}$, these events can be vetoed. The energy spectrum of the ${K_L}^0$ is sharply falling with energy reducing the probability of producing muons, with energies $>$0.5~GeV, which can be identified and reconstructed in ICAL. Finally the ${K_L}^0$ interactions will contribute to events originating on the periphery of the detector ($\lambda_{Had}$ (Fe) $\sim$ 17~cm) which would lead to a modest reduction in the fiducial volume. A realistic estimate of such a background would require a simulation which will be dealt with separately. However it does seem that this will not be prohibitively large as to vitiate the use of a ICAL-CMV at shallow depth.

Since the CMV detector was assembled from old plastic scintillators that were available in our laboratory, the measured 99.98\% veto efficiency is very encouraging. A better designed detector, such as the Mu2e CRV detector \citep{cdr_mu2e}, could possibly give a much better efficiency. A proof-of-principle CMV is being planned for the 80~ton mini-ICAL (4~m $\times$ 4~m $\times$ 10 layers with a centrally located 2~m $\times$ 2~m Resistive Plate Chamber in each layer) that is being built at IICHEP, Madurai. 

\section{On a possible CMV for a shallow depth ICAL and proof of principle demonstration in mini-ICAL}

As mentioned in the introduction the motivation for this work was to explore the possibility of placing an atmospheric neutrino detector at a shallow depth rather than in a deep underground location. The advantages of locating a large detector such as ICAL at a shallow depth of 100~m, are obvious. While the number of underground locations that satisfy the criteria of being ecologically friendly, seismically quiet, good rock overburden of about 1000~m on all sides is small, if one relaxes the last criterion this number increases dramatically. At first sight such a possibility does seem outlandish, given the large cosmic muon background at the shallow depth of about 100~m at ICAL of about 3$\times10^{8}$ /day \citep{Bogdanova} and the small atmospheric neutrino event rate of about 3 per day in the same detector. On the other hand if one were to build a cosmic muon veto detector which could detect muons with > 99.99$\%$ efficiency, the fraction that would escape detection would be of the same order or similar to that which survives after traversing a rock cover of about 1~km.

Given the size of ICAL a large CMV would be required. In view of the possible use of ICAL with auxiliary detectors on the walls of the cavern, in searches for primordial magnetic monopoles \citep{nitali2015} and anomalous KGF events \citep{narasimham2004}, the CMV could double up for this purpose as well. In such a scenario about 2-3 layers of a plastic scintillator could be used covering the top and 4 sides of an enclosed volume with overall dimensions of $\sim$ 50~m $\times$ 25~m $\times$ 16~m. During the course of the assembly of the CMV detector described in Section 2, we became aware of a similar, but more advanced effort in connection with the Mu2e experiment at Fermilab \citep{cdr_mu2e}. The scintillator wall could be basically of the type planned for Mu2e.

The possibility of constructing a CMV for a shallow depth ICAL, needs to be validated. The advantage of a shallow depth is the low background from the primary hadronic and electromagnetic components in cosmic rays. Such a proof-of-principle CMV is being planned to be built surrounding the mini-ICAL detector at IICHEP, Madurai. As the mini-ICAL detector, with 10 layers of RPC, has a magnet of dimensions $\sim$4~m $\times$ 4~m $\times$ 1.6~m the CMV of dimensions $\sim$5~m $\times$ 5~m $\times$ 2~m is needed. As each layer will be populated with 2 RPCs of dimensions $\sim$2~m $\times$ 2~m the muon trigger rate is expected to be about 500~Hz leading to about 4.3 $\times10{^7}$ triggers/day.  A CMV detector with an inefficiency of $10^{-6}$ for cosmic muons could be measured to an accuracy of 10 \% within a couple of days. The muon identification would be clean due to the tracking in the RPC detectors. Neutron and $\gamma$-ray induced events in mini-ICAL could be estimated from the simultaneous absence of hits in the CMV detector. This would be useful in assessing whether a shallow depth ICAL is feasible or not. If the answer is positive one could envisage much larger iron calorimeter detectors for possible use in an atmospheric neutrino experiment for using the matter effect for neutrino science as well as applications such as earth tomography.  

\section{Summary}

In summary, the results of an experiment to measure the efficiency of a cosmic muon veto detector of about $\sim$ 1~m $\times$ 1~m $\times$ 0.3~m dimensions are reported with a view to assess the feasibility of building a larger version for a shallow depth ICAL detector. Efficiency maps of the plastic detectors used in the veto detector using a RPC stack for cosmic muons are also reported. The measured veto efficiency of (99.978 $\pm$ 0.003)\% with 3 layers of plastic at the top appears to be sufficiently encouraging as to justify building a larger $"$proof-of-principle$"$ veto detector covering the mini-ICAL detector which will measure cosmic muons. Its success will determine whether or not one can build the much larger ICAL detector for atmospheric neutrinos at a shallow depth. 


\acknowledgments

We would like thank our INO colleagues at TIFR and project student Neha Lad for help during the setting up of the experiment. We would also like to thank B.S. Acharya, N.K. Mondal and D. Indumathi for their critical reading of the manuscript. One of the authors (VMD) would like to thank M.V.N. Murthy for the discussions that prompted this investigation.
\bibliographystyle{JHEP}
\bibliography{muon_veto}
\end{document}